# SYSTEMS-THEORETIC HAZARD ANALYSIS OF DIGITAL HUMAN-SYSTEM INTERFACE RELEVANT TO REACTOR TRIP

**Edward Chen and Nam T. Dinh**
Department of Nuclear Engineering, North Carolina State University
Raleigh, NC, 27695, USA
Echen2@ncsu.edu, Ntdinh@ncsu.edu

**Han Bao and Hongbin Zhang**
Idaho National Laboratory
2525 Fremont Ave, Idaho Falls, ID, 83402, USA
Han.Bao@inl.gov, Hongbin.Zhang@inl.gov

**Tate Shorthill**
Department of Mechanical Engineering and Materials Science, University of Pittsburgh
3700 O'Hara Street, Pittsburgh, PA 15261
ths60@pitt.edu

## ABSTRACT

Human-system interface is one of the key advanced design features applied to modern digital instrumentation and control systems of nuclear power plants. The conventional design is based on a compact workstation-based system within the control room. The compact workstation provides both a strategic operating environment while also a convenient display for plant status information necessary to the operator. The control environment is further enhanced through display panels, visual and auditory alarms, and procedure systems. However, just like the legacy control, the HSI should incorporate diversity to demonstrate sufficient defense-in-depth protection against common cause failures of the safety system. Furthermore, the vulnerability of the HSI is affected by a plethora of factors, such as human error, cyberattacks, software common cause failures, etc., that complicate the design and analysis. Therefore, this work aims to identify and evaluate existing system vulnerabilities to support the licensing, deployment, and operation of HSI designs, especially the functions that are relevant to a reactor trip. We performed a systematic hazard analysis to investigate potential vulnerabilities within the HSI design using the novel redundancy-guided systems-theoretic hazard analysis. This method was developed and demonstrated by Idaho National Laboratory under a project initiated by the Risk-Informed Systems Analysis Pathway of the U.S. Department of Energy's Light Water Reactor Sustainability Program. The goal of the project is to develop a strong technical basis for risk assessment strategies to support effective, reliable, and licensable digital instrumentation and control technologies.

*Key Words*: Hazard analysis, digital control, common cause failures, human-system interface

## 1  INTRODUCTION

In recent years, analog controls in nuclear power plants (NPP) are increasingly replaced with digital hardware. With digitalization, issues within the plant can be immediately detected with real-time network monitoring, boosting the overall safety and reliability of the system (e.g., Asset Performance Manager [1]). Furthermore, by replacing analog hardware with distributed programmable logic controllers (PLCs) or distributed control systems (DCS), each subsystem can act more independently and can respond to the output of the other, enhancing the overall automation of the reactor [2]. New digital instrumentation and control (DI&C) systems implemented in the Korean OPR-1000 reactors [3] have been added to the design of the APR1400 reactor [4]. However, increased digitalization adds a considerable amount of risk to all operation levels.

One particular module of interest is the human-system interface (HSI). In digitalizing the main control room, flat panel displays, soft control modules, and online direct monitoring systems (e.g., +SPADES) replace analog meters, dials, and control systems. In addition, intermediate safety critical values can also be determined in advance in real time and reduce the cognitive load on operators, for example metrics for inadequate core cooling (ICC). The growing dependency on digital information systems, especially those relevant to safe operation, also raises the question of the reliability and accountability of received data. Software failures modes and common cause failures (CCFs), especially in the HSI, can considerably reduce the operability and safety of NPPs. For example, in the APR-1400 reactor protection system, there are 16 identical processors that compose the local coincidence logic unit [4]. One failure could lead to the common failure of all processors, impacting reactor trip functionality. This makes the identification of CCFs, especially in nondiverse safety critical systems, incredibly vital to the risk assessment of new control systems.

Unlike analog components, digital design is relatively new and constantly evolving, making the risk assessment of components difficult. Furthermore, the efficacy of commercial-off-the-shelf software and hardware for nuclear applications has not been adequately analyzed and may not be compliant with existing industry standards (e.g., IEEE 308-2020 [5], NQA-1 [6]). These, in combination with the increased interdependency of digital components, result in a lack of adequate statistical data on different system and equipment failure rates, further complicating the risk assessment and hazard analysis of DI&C systems.

In addition, conventional risk assessment and hazard analysis methods for analog systems, such as fault tree analysis (FTA), failure mode and effects analysis (FMEA), and probability risk assessment, are insufficient for digital systems. In separate analyses on the root causes of software failures, the failures were primarily due to an inadequacy of design and requirement constraints as opposed to random failures [7, 8]. Furthermore, a report from the Electric Power Research Institute on estimating failure rates in highly reliability digital systems [9] concluded that traditional methods to determine software failures, especially CCFs, were insufficient and revised methods were required.

With context defined, this paper thus presents a case study to identify potential software failure modes and CCF within the digital HSI. We analyze the underlying redundant hardware design and outline possible software failure modes.

## 2  TECHNICAL BACKGROUND

In this section, we conduct a literature review on existing hazard analysis methods for digital software systems. Software hazards are notoriously difficult to categorize into disparate failed states. Inappropriate actions propagated by the governing control algorithm could be considered hazardous in the wrong timing or context. Furthermore, even if no individual software module fails, it is still possible that the overall system fails to achieve the primary directive through unanticipated software interactions. In that respect, all digital controls signals could be considered hazardous, especially when applied inappropriately or inadequately [7]. While there is no general industry consensus regarding software failure modes and probability, different methods have been proposed to address these issues.

The most popular and informative method to conduct hazard analysis is static FTA [2]. By first identifying the top event of concern, different causes of system-level failures can be discovered. Fault trees (FTs) have been used extensively in multiple industries due their simplicity in capturing complex failure scenarios. However, static FTs are severely limited in modeling component interactions, timing, and the physical plant process, all of which all relevant to identifying software inadequacies [10].

Another widely used industry method is software failure modes and effects analysis (SFMEA). Like hardware FMEA, it also seeks to determine system-level effects whenever any single module fails in a certain way [10]. SFMEA specifically excels at identifying high-level speculative failure modes in a software system and can be used to design effective methods to advert them. However, detailed SFMEA is incredibly complex due to a range of issues, ultimately making the method inadequate for low-level software hazard analysis. Major SFMEA issues involve an inability to assess vendor software proprietary source code and the unpredictability of system consequences [10]. The latter problem is due to the highly interdependent relationship between variables and modules, making it incredibly difficult to definitively predict the consequences of failed states. Minor issues with the method include determining the extended effects of software variables impacting multiple modules, the applicability of specific failure modes under different scenarios, and the legitimacy of variables based on time dependency [10].

Systems-theoretic process analysis (STPA) [7] was developed to address these issues by qualitatively identifying all potentially hazardous control actions (CA) and component interactions that could lead to a defined loss. Instead of identifying all possible low-level component interactions and potential failure modes, STPA generalizes failures into four categories of unsafe control actions (UCAs): (1) CA are missing when needed, (2) CA are applied when not needed, (3) CA are applied too early, too late, or in the wrong order, and (4) CA are applied too short or too long. STPA can also be applied to low-level control modules and can utilize hierarchal recursive approaches similar to FTs. While effective, STPA does not clearly outline methods to identify CCF or quantify the probabilities of identified UCAs.

In response, Hazard and Consequence Analysis for Digital Systems (HAZCADS) [12] was developed for a more comprehensive system analysis. HAZCADS is a combination of two existing methods, FTA and STPA, and can be roughly separated into two stages. In the first stage, STPA is applied to identify potential UCAs [7]. In the second stage, the identified relevant UCAs to a particular top event are then integrated into an FT where minimal cut sets can be deduced to find potential single points of failure (SPOF) [12]. However, while HAZCADS identifies potential software faults and integrates them into the hardware FT, the process lacks the clarity required to analyze multilayered redundant systems.

Building upon the HAZCADS methodology, Idaho National Laboratory (INL) created redundancy-guided systems-theoretic hazard analysis (RESHA) [13-15] to further address and clarify issues and potential sCCF across multiple redundant components. INL developed and demonstrated RESHA under a project initiated by the Risk-Informed Systems Analysis Pathway of the U.S. Department of Energy's Light Water Reactor Sustainability Program [16, 17]. The aim of the project is to develop a strong technical basis for risk assessment strategies to support effective, reliable, and licensable DI&C technologies.

Typically, there are four types of CCFs, categorized by dependency of systems, structures, and components (SSCs) [13]. In a Type 1 CCF, two or more SSCs depend on the same controller or resource. A failure of the shared resource subsequently causes both SSCs to fail accordingly. Type 2 and Type 3 CCFs are similar to each other, both depending on a shared resource; however, in Type 3 the cause of failure is to a shared resource external to the controller. Finally, a Type 4 CCF is where two separate controllers experience a failure due to a shared common design or location. There are several potential failures causes for CCF occurrence in a system. The first is a random hardware failure, such as a soldering disconnect or a capacitor wearing out. These can typically be characterized with standard stochastic failure probabilities determined during the quality assurance phase of development. However, more latent causes, such as design defects, are systematic in nature and cannot be completely removed through testing. Defects can manifest when the constraints on requirements are not properly defined during the initial development phase [7]. While unit testing can remove some of these defects, probing every module to the detail required within a software to discover all defects is infeasible, making design defects a significant challenge. In addition, the defects can result in unanticipated scenarios during plant operation, leading to unaccounted for losses. Another cause are environmental hazards, such as earthquakes causing a CCF of redundant SSCs (e.g., servers) located on the same floor. Human error during operation or testing is another potential source of CCFs.

## 3    CASE STUDY

This section conducts a hazards analysis via RESHA on an advanced HSI relevant to reactor trip safety developed directly from the Advanced Pressurized Reactor (APR) 1400 HSI. From documentation [4], the HSI of the APR1400 consists primarily of four redundant information retrieval systems for the operator. During nominal reactor operation, the Qualified Indication and Alarm System-Non-Safety (QIAS-N) receives analog and digital signals from both safety and non-safety-related plant components. Under both nominal and safety relevant scenarios, the Qualified Indication and Alarm System-Safety (QIAS-P) acts as both a continuous source of accident monitoring information as well as a backup operator display module to the QIAS-N system [4]. The primary role of the QIAS-P system (Appendix A) is to provide an unambiguous indication of ICC as well as advanced warning of the approach towards it [4]. Both QIAS systems are implemented using a PLC-based control platform. To introduce HSI diversity, an additional Information Processing System (IPS) also collects relevant sensor information and plant states [4]. The IPS is implemented under a DCS-based platform, which is fundamentally different than the QIAS system. In speculated software failures, either common cause or not of both the QIAS and IPS digital control systems, a last resort all analog Distributed Instrumentation System (DIS) is available that monitors the same key accident monitoring instrumentation variables as the QIAS-P. This diversity in the HSI system provides multiple information routes to the operator to safely shutdown the reactor either under nominal, anticipated operational occurrences, or speculated software CCF (sCCF). To demonstrate the 7-step RESHA process, we analyzed a representation of the QIAS-P system from the APR1400.

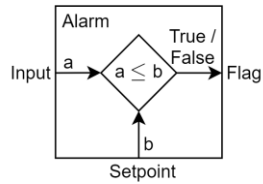

**Figure 1. Basic comparator diagram for alarm components.**

### 3.1.1   Assumption 1: all set points are manually assigned and alarms are comparators

From documentation, there is explicit mention of alarms utilized in the QIAS-P system; however, the internal architecture of the alarm is not specified. Therefore, we assumed the most basic version of an alarm (see Fig. 1), that is, a simple comparator with a single set point trigger. The alarm set points are assumed to be manually assigned by the operator as no details are available. This increases the overall number of possible software failures.

### 3.1.2   Assumption 2: internal architecture is approximated

We pieced together the development of the detailed hardware architecture from multiple documentation and published sources on the APR1400 [4, 19-25], as an exact and verified diagram is proprietary. Therefore, the constructed diagrams included in this paper are approximate or fictious representations based on the described connections and functionality. Some of the failures included in the hardware FT may not be reflected in the real system. In addition, results derived from the RESHA method are meant for the application demonstration and are not meant to be real guidelines for the HSI system of the APR1400.

### 3.1.3 Assumption 3: sensors act and fail in unison

The QIAS-P system receives 61 core exit thermocouples (CET) and 34 heated joule thermocouple (HJTC) sensors directly from the core. These sensors are assumed to operate as a single unit, and a failure of one sensor results in the failure of the entire sensor array. This is to avoid partial failure modes and unsupported speculation on architecture response as the behavior and handling of the sensor information received by the QIAS-P system lacks detailed documentation.

### 3.1.4 Assumption 4: distributed network architecture is applied

From the APR1400 documentation [4], the QIAS systems are implemented with a network of PLC that communicate to other modules via a safety data network (SDN). However, it is not explicitly mentioned how the QIAS systems communicate with the SDN. Typical industrial networks operate via distributed node-based communication (e.g., Supervisory Control and Data Acquisition [17]) as opposed to centralized server-based communication. The QIAS-P representation will be assumed to be node-based as well. This reduces the number of communication layers between the operator and information system that can fail.

### 3.1.5 Assumption 5: hardware design failures ignored

This assumption applies to the hardware design choices made during the development of the digital and analog proportions of the QIAS-P. It includes, but is not limited to, which power supply to use, integrated chip sets, connectors, circuits, etc. Due to the unavailability of a detailed and proprietary hardware schematic, the hardware design failure branch is excluded from the FT and will not be shown.

### 3.1.6 Assumption 6: QIAS-P redundant division architecture is not diverse

While typical development practices are to introduce diverse design in redundant divisions of the QIAS-P to prevent CCFs, it is not explicitly mentioned in the documentation that diverse design was considered. Instead, system diversity is accomplished by designing the QIAS-N, IPS, and DIS differently. While it is possible that hardware and software diversity was considered in the implantation of the QIAS-P, in this paper, we assumed that the two redundant divisions are identical.

### 3.1.7 Assumption 7: operator decision models are noncomplex based solely on data

Due to the complex nature of human decision making, misinformation propagated by the QIAS-P system to the operator does not guarantee failure. In fact, an additional separate human reliability analysis is required to survey the impact on falsified information on the operator decision model. However, in this case study, we assumed that failures by the QIAS-P system to communicate correct information or trigger the correct alarm will cause a failure in the operator decision model.

## 3.2 Stage 1: Create a Detailed Hardware Representation of the Digital System of Interest

When examining a digital system of interest, a physical and literal representation of the underlying hardware is required. Specifically, information flow and feedback between separate modules should be identified. We developed the hardware representation for a redundant division of the QIAS-P, and it can be seen in Appendix A.

The QIAS-P monitoring system is composed of two redundant divisions (A&B) isolated both physically in location and in communication from each other. Within each division, there is a digital processing module (PM) and an analog retrieval module (AM). The two modules operate sequentially to retrieve, process, and check values and alarms sent to the operator information terminal in the MCR.

The primary purpose of the PM is to check and determine intermediate values for an operator display and use in the internal alarm system. From documentation, the PM is composed of five primary parameter calculators and alarms, namely the HJTC, reactor vessel level (RVL), reactor coolant saturation margin (RCSM), ICC, and the CET temperature. The PM also includes an HJTC heater power controller, a maintenance/interface test panel (MTP/ITP), and various secondary parameter calculators and alarms. One of the subroutines of the PM is to modulate the HJTC reference power level based on a power setpoint. This reference level is used to calculate the heat differential across the HJTC sensing junction [24]. A manual override of the HJTC power controller is also available to switch control to the DIS in a postulated sCCF.

The primary purpose of the AM is to retrieve signals from the various sensors and subsystems and either convert them from analog to digital (via analog to digital converter) or to repeat or convert the signal (via signal conditioners). In total, the AM module receives 32 HJTC and 61 CET sensor values directly from the core. In addition, it also receives the hot and cold leg temperatures, pressurizer pressure, and reactor vessel head as well as Type A, B, and C variables from the SDN or auxiliary processing cabinet. Lastly, to ensure sensor diversity, half of the analog HJTC and CET values are also routed to the DIS.

### 3.3 Stage 2: Develop a Hardware FT for the Top Event of Interest in the Digital System

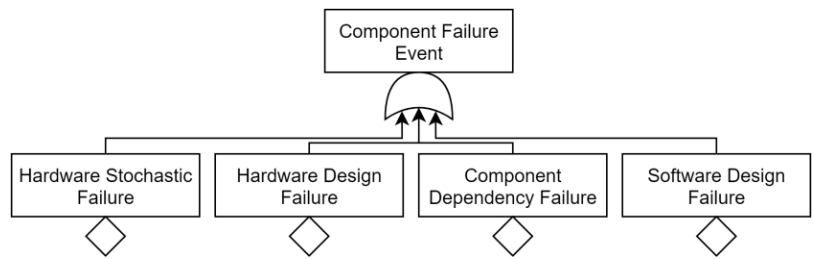

**Figure 2. Generic component failure and recommended failure branches.**

Based on detailed hardware representation, FTs can be developed to identify potential hardware and dependency failures. It is recommended that, for each event, four different failure branches are included (see Fig. 2), namely the hardware stochastic failure, the hardware design failure, the dependency failure, and an unresolved software design failure branch. The first two failure branches describe a normal wear of hardware as well as potential failures in hardware design constraints. The dependency failure branch is allocated for when dependent signals from other components are missing, causing a failure. Failure probability quantification is not required at this stage.

By using the detailed hardware model developed in Stage 1 (see Appendix A) and the guidelines from Fig. 2, the full hardware FT can be developed. The top event in the FT is defined as the "Operator fails to initiate reactor trip causing reactor damage." Based on this top event, the full FT includes 41 hardware stochastic failure basic events and 33 component dependency failure branches. There are also 26 unresolved software design failure branches. A system-level hardware FTs is developed and shown in Appendix B as an example. System-level redundancies (i.e., IPS, QIAS-N, and DIS) are identified but unanalyzed.

### 3.4 Stage 3: Determine UCAs/UIFs Based on a Redundancy-Guided Application of STPA

In this stage, an expanded version of STPA is applied to identify potential software failures. In the original STPA, component interactions with other components leading to hazards are UCAs. Specifically, a UCA is only possible when a controller gives a command and receives feedback information from the dependent component, that is a closed control loop (see Fig. 3). However, unsafe information flows (UIFs) leading to unanticipated program faults that are not necessarily caused by CA are also possible. In the second scenario, a component provides fabricated or false data to another dependent component that it relies on for decision making. The dependent component processes the false data and outputs false signals, leading to a hazardous state (see Fig. 3). This results in two possible causes of software failures, either UCAs or an UIF. Thus, we applied the revised STPA to identify these two types of events.

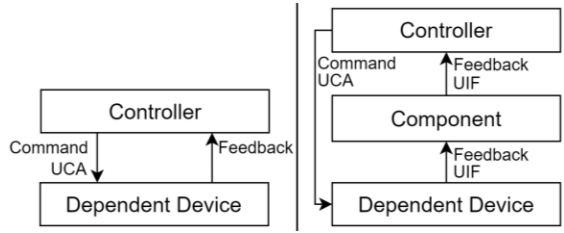

**Figure 3. (Left) Original STPA control diagram. (Right) Revised STPA diagram for RESHA.**

The first step in STPA is to define the purpose of the analysis and to identify losses, hazards, and the analysis scope. For the QIAS-P system, the losses and hazards are defined in Table I. Defined losses are typical to any operational circumstance, while defined hazards are specific to QIAS-P. Due to a lack of controllers within the monitoring system, the primary impacted "component" related to reactor trip safety is the human operator. Therefore, when the QIAS-P system provides misleading information or spurious alarms to the operator, one of the defined losses is possible. For example, a false positive alarm (H-3) could mislead the operator to unnecessarily trip the reactor, leading to a loss of plant availability (L-4).

After identifying losses and hazards, the scope of this analysis is defined. In this paper, the analysis is limited exclusively to all components and redundant systems within QIAS-P. Signals conveyed to exterior components, such as indication and alarm values to the QIAS-N, DIS, and IPS, are not covered in the analysis. The FT in Appendix B provides a good representation of the analysis scope. The specific purpose is to identify UCAs or information flows as well as causal events that could lead to the hazardous states defined in Table I.

**Table I. Defined losses and hazards for QIAS-P HSI.**

| Losses | Hazards |
| --- | --- |
| L-1 Damage to reactor or key reactor components | H-1 QIAS-P false positive indication |
| L-2 Damage to operational equipment | H-2 QIAS-P false negative indication |
| L-3 Damage to monitoring & control hardware | H-3 QIAS-P false positive alarm |
| L-4 Loss of plant availability | H-4 QIAS-P false negative alarm |
| L-5 Generic: loss of life, environmental contamination | |

After defining the purpose of the analysis, the second step is to model the control diagram based on the detailed hardware representation. Not all components in the detailed hardware diagram are represented in the control diagram; instead, only components that can alter the state of other components are required. Here, the levels of redundancy (i.e., system, division, module) as well as the information flow between the various components should be clearly outlined. In Fig. 4, the different levels of redundancy and information flow between a condensed version of QIAS-P can be seen. The third step in STPA is to identify UCAs based on the control model previously developed. UIF is also identified in this stage and follows the same guiding principles.

From the four categories of UCAs, we defined six subcategories to better separate different scenarios (Table II). Both UCAs and UIFs put the system only in hazardous states and require a scenario to cause a loss. From the control diagram (see Fig. 4), three UCAs were identified related to the HJTC controller, 15 UIFs were identified related to the digital parameter calculators, and 10 UIFs were identified related to the digital alarm systems. Examining the HJTC controller, the UCAs involved are the controller fails to provide a power reference level to the HJTC sensors when needed (Type A UCA) and the controller provides a reference level but is either too high (Type F UCA) or too low (Type G UCA). For parameter calculators, the prominent UIFs are the calculator fails to output a value when needed (e.g., zero, null, inf) (Type A UIF) and the value calculated is either too high (Type F UIF) or too low (Type G UIF). For alarms, based on the

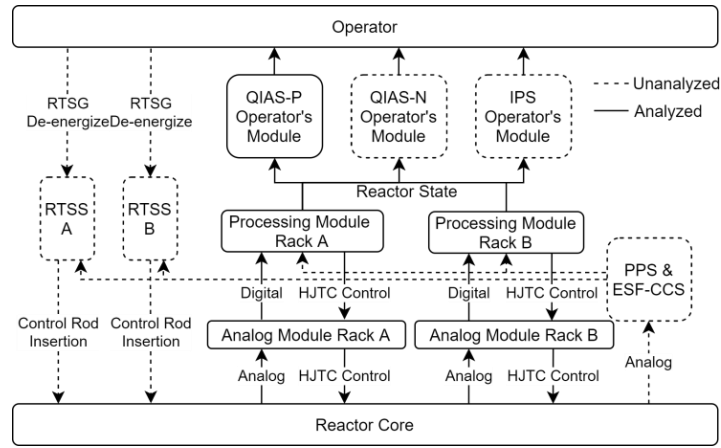

**Figure 4. Condensed qualified indication and alarm control system-safety (QIAS-P) control diagram.**

Assumption 1, there are only two UIFs per alarm, either it fails to trigger when needed (Type A UIF) or triggers when not needed (Type B UIF).

**Table II. Reclassified UCAs and information flows from STPA**

| STPA-Defined Control Action | RESHA-Defined Unsafe Control Action or Information Flow |
|---|---|
| CA is missing when needed | (A) CA is missing when needed |
| CA is provided when not needed | (B) CA is provided when not needed |
| CA is provided too early or too late, out of order | (C) CA is provided too early |
| | (D) CA is provided too late |
| | (E) CA is applied in the wrong order |
| CA is applied too long or stopped too soon | (F) CA is applied too long or too much |
| | (G) CA is stopped too early or applied too little |

### 3.5 Stage 4: Construct an Integrated FT by Adding Applicable UCAs/UIFs as Basic Events

In this stage, based on the selected top event of the FT, relevant unsafe software actions are added into the hardware FT under the software design failure branch. Multiple unsafe actions may appear in more than one location in the FTs due to the interdependency of system components.

For the HJTC power controller, the relevant context event is when the controller fails to regulate the HJTC reference power correctly. The three UCAs identified in the previous stage are all relevant to this top event (see Fig. 5). Incorrect reference power levels cause the sensed signal to be either higher or lower than the true value, further misleading other components in the system. For calculators, the generic top event is when it fails to output the proper values when all dependent values are correct. This is purely a software problem directly related to the mathematical "model" used in calculation and applies to the three UIFs identified.

## 3.6 Stage 5: Identify Software CCFs from Duplicate UCAs/UIFs Within Integrated FT

Duplicate unsafe software actions that appear in the integrated FTs are considered sCCF. Depending on the level of redundancy, different CCFs are possible, as defined in Section 2. A partial integrated FT with division and system-level sCCFs can be seen in Fig. 6.

At the QIAS-P system level, from Assumption 6, 28 system-level sCCFs were identified affecting all digital components in both redundant systems. Regarding the HJTC controller, failure in the controller software would render simultaneous failures in HJTC sensors, regardless of physical signal isolation. Accordingly, for the three UCAs identified, three Type 4 sCCF are possible for the HJTC power controller. A similar approach is taken for the 25 UIFs related to the five parameter calculators and alarms, resulting in 25 Type 4 sCCFs.

At the QIAS-P division level, 15 division-level sCCFs were identified affecting some or all digital components. From analysis, it was determined UCAs related to the HJTC power controller affect the most components. Due to the strong dependency of all components in the QIAS-P system on the HJTC sensor readings, UCA Type A, E, and F by the controller would cause misleading outputs on four of five calculators and alarms. To clarify, the HJTC reference power controlled by the HJTC controller dictates the temperature readings recorded by the sensor. Incorrect reference power levels would cause the sensor to return higher or lower readings. However, the sensor itself is not broken and operates normally. Affected modules include the HJTC, ICC, RVL, and RCSM calculators and alarms (8 total). In total, three Type 2 sCCFs associated with the three UCAs are possible by the controller. For the parameter calculators, the afflicted modules with possible sCCF are the HJTC, CET, RVL, and RCSM. Due to the interdependence on these calculators, UIFs from them will cause at most the failures of two other components. For example, the CET calculator output is used in the ICC alarm and the CET temperature alarm. Type A, E, or F from the CET calculator would cause the corresponding alarms to fail. In total, there are 12 Type 2 sCCF possible related to the four afflicted calculators each with 3 UIFs.

## 3.7 Stage 6: Determine the Minimal Cut Sets to Discover Potential SPOFs

Once all relevant CCFs are identified, whether hardware or software, the minimal cut sets can be determined to discover potential single points of failures within the system of interest. For low levels of architectural diversity, it is likely that many first order cut sets as failures in design constraints are repeated in multiple redundant control divisions.

In this case study, the top event is "Operator fails to initiate reactor trip causing reactor damage." As this study is limited to the QIAS-P system, redundancy provided by the IPS, DPS, and QIAS-N is not considered. With those considerations, multiple first order cut sets were discovered that can trigger the top event. In total, 43 software first order cut sets were discovered. Only five sets are listed in the table as an example (Table 3).

Using Cut Set #1 as an example, due to similar designs, a latent software failure in the HJTC controller would affect both division A & B. If this causes the controller to provide power reference levels lower that the set reference level (UCA Type F), the HJTC sensors would have a higher than anticipated reading. The HJTC, RVL, and RCSM calculators that depend on this value would then also report higher than normal readings. Depending on the severity of the UCA, this could cause the HJTC temperature, ICC, RVL, and RCS alarm to trigger, creating H-2 and H-4 hazards. Based on Assumption 7, the operator would immediately trip the reactor, resulting in an L-4 loss.

## 3.8 Stage 7: Identify and Provide Guidance to Eliminate Latent Faults or Triggers of CCFs

The primary causes of CCF in the system can be traced back to the lack of software and hardware diversity by the two QIAS-P redundant divisions. This argument is supported in the documentation as it is neither explicitly nor implicitly implied that design diversity was considered [4]. The justification for the original designers was that the IPS and DIS would act as diverse backups to the QIAS-P system. Nevertheless, as the first primary source of information, the operator, diversity among QIAS-P divisions should be implemented.

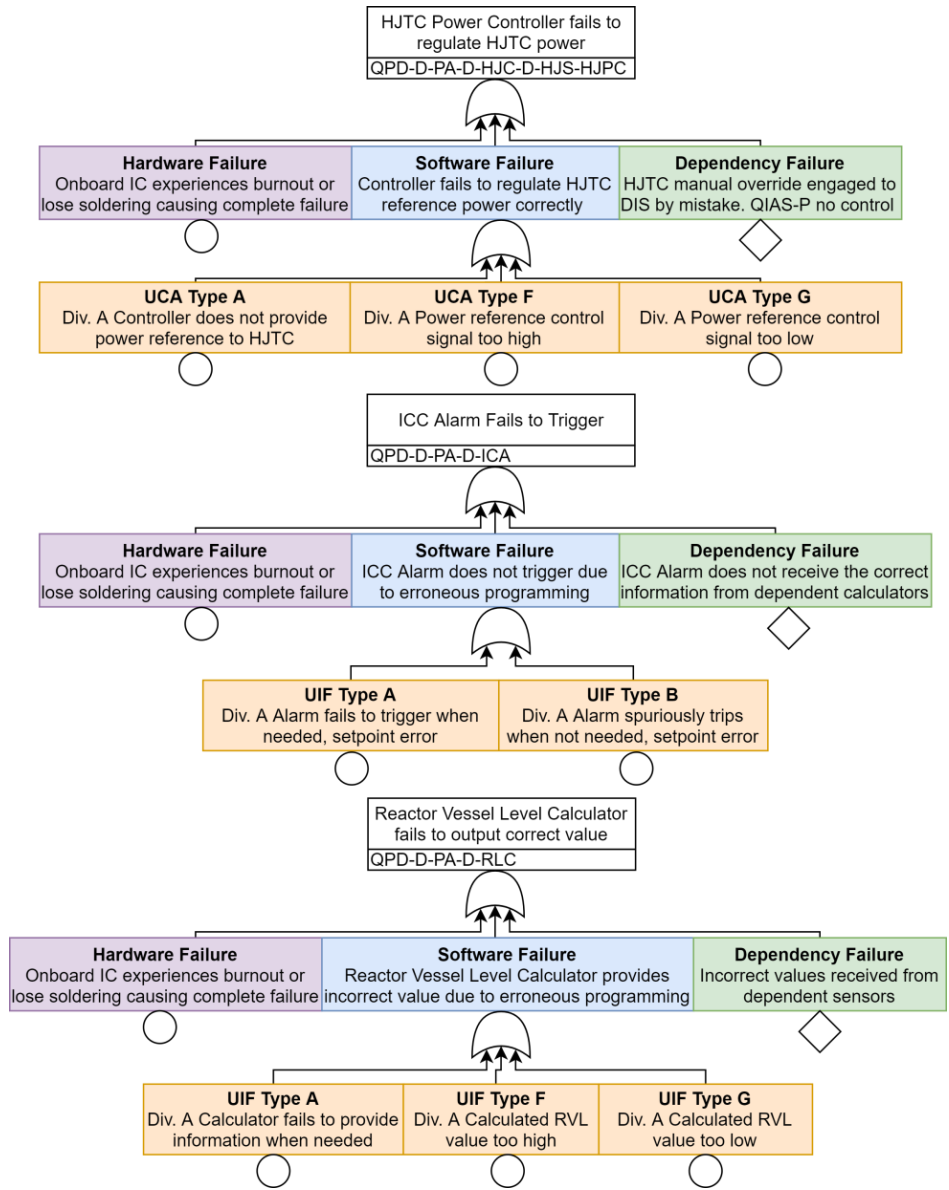

Figure 5. HJTC power controller (Top), ICC alarm (Middle), and RVL calculator (Bottom) failures and associated UCAs/UIFs.

Table III. First order software cut sets

| # | Cut Set / Basic Event Description |
|---|---|
| 1 | Division A&B HJTC controllers fail to provide correct power reference to HJTC sensors |
| 2 | Division A&B HJTC calculators provide incorrect/misleading parameter values |
| 3 | Division A&B CET calculators provide incorrect/misleading parameter values |
| 4 | Division A&B RVL calculator provide incorrect/misleading parameter values |
| 5 | Division A&B RCSM calculators provide incorrect/misleading parameter values |

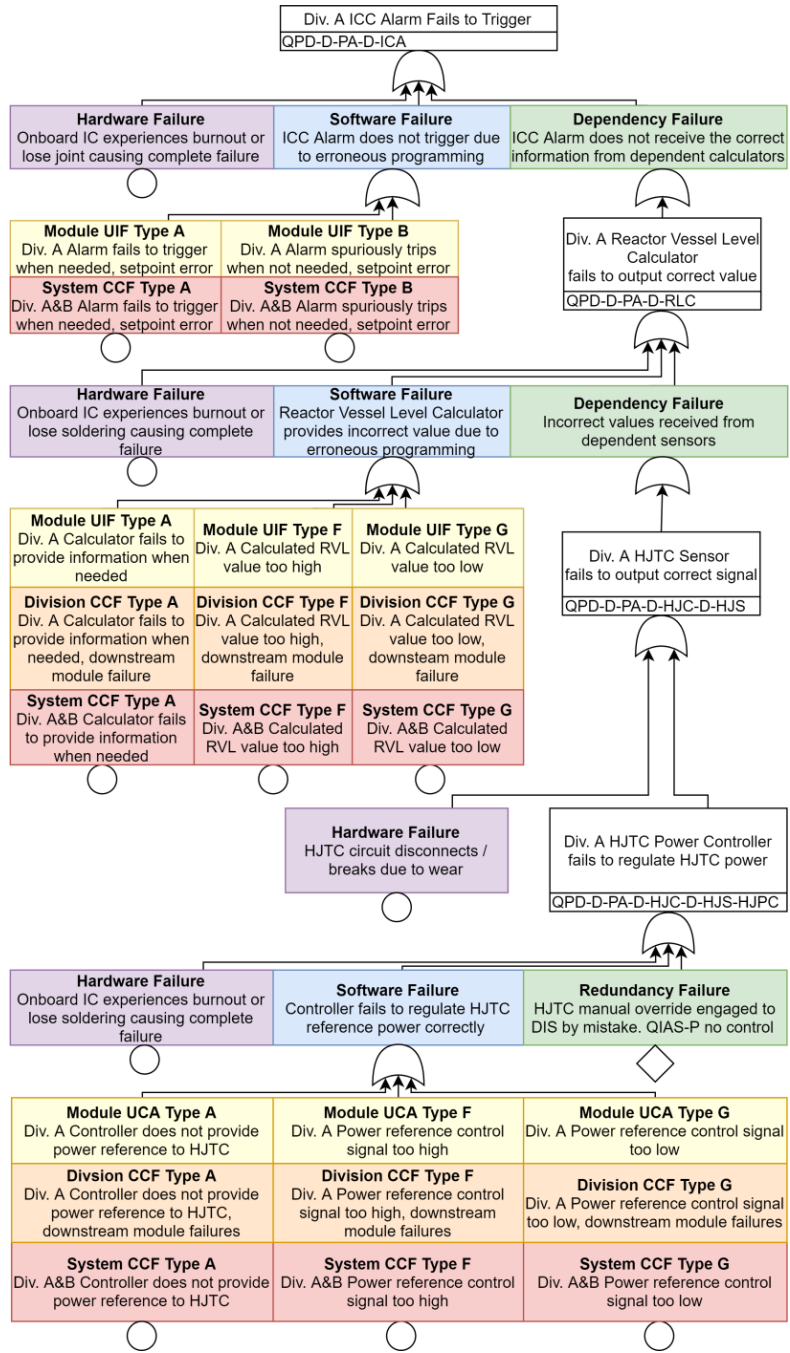

Figure 6. Partial integrated FT with Division A.

The secondary causes of CCF in the system are related to the strong interdependency of software modules within a specific division. For example, almost all components in the QIAS-P require readings from the HJTC sensors. A failure of the HJTC controller would therefore result in a systemwide failure.

Finally, different latent software failures have different possible sources. Type A and B UCA/UIFs could be caused by failures in the programming stage, where either the output variable is unassigned after calculation or the setpoint variable is under or over the ideal limit. Type F and G UCA/UIFs are directly caused by the inappropriate boundary conditions of the module or an incorrect process model creating unreliable output calculations.

## 4   CONCLUSIONS

This paper applied RESHA to identify potential hazards and CCFs for a representation of the APR1400 HSI. Specifically, we analyzed the QIAS-P information system and identified sCCFs relevant to reactor trip safety. Due to the assumed lack of diversity in the architecture, the system is highly susceptible to Type 4 CCFs, where failures in either the design or development would cause failures in all redundancies. Furthermore, due to the strong reliance on the HJTC sensors, especially the controller, incorrect or missing reference power levels can lead to a Type 2 sCCF potentially contributing to a SPOF, resulting in L-1 through L-4 losses depending on the triggered UCA. This analysis demonstrates the capability of RESHA to qualitatively identify hazards at all relevant redundancy levels.

In future work, the specific failure probability of each basic event will be identified, and the probability of relevant top events calculated using Bayesian Belief networks described in BAHAMAS [14].

## 5   ACKNOWLEDGMENTS

This submitted manuscript was authored by a contractor of the U.S. Government under DOE Contract No. DE-AC07-05ID14517. Accordingly, the U.S. Government retains and the publisher, by accepting the article for publication, acknowledges that the U.S. Government retains a nonexclusive, paid-up, irrevocable, worldwide license to publish or reproduce the published form of this manuscript, or allow others to do so, for U.S. Government purposes. This information was prepared as an account of work sponsored by an agency of the U.S. Government. Neither the U.S. Government nor any agency thereof, nor any of their employees, makes any warranty, express or implied, or assumes any legal liability or responsibility for the accuracy, completeness, or usefulness of any information, apparatus, product, or process disclosed, or represents that its use would not infringe privately owned rights. References herein to any specific commercial product, process, or service by trade name, trademark, manufacturer, or otherwise, does not necessarily constitute or imply its endorsement, recommendation, or favoring by the U.S. Government or any agency thereof. The views and opinions of authors expressed herein do not necessarily state or reflect those of the U.S. Government or any agency thereof.

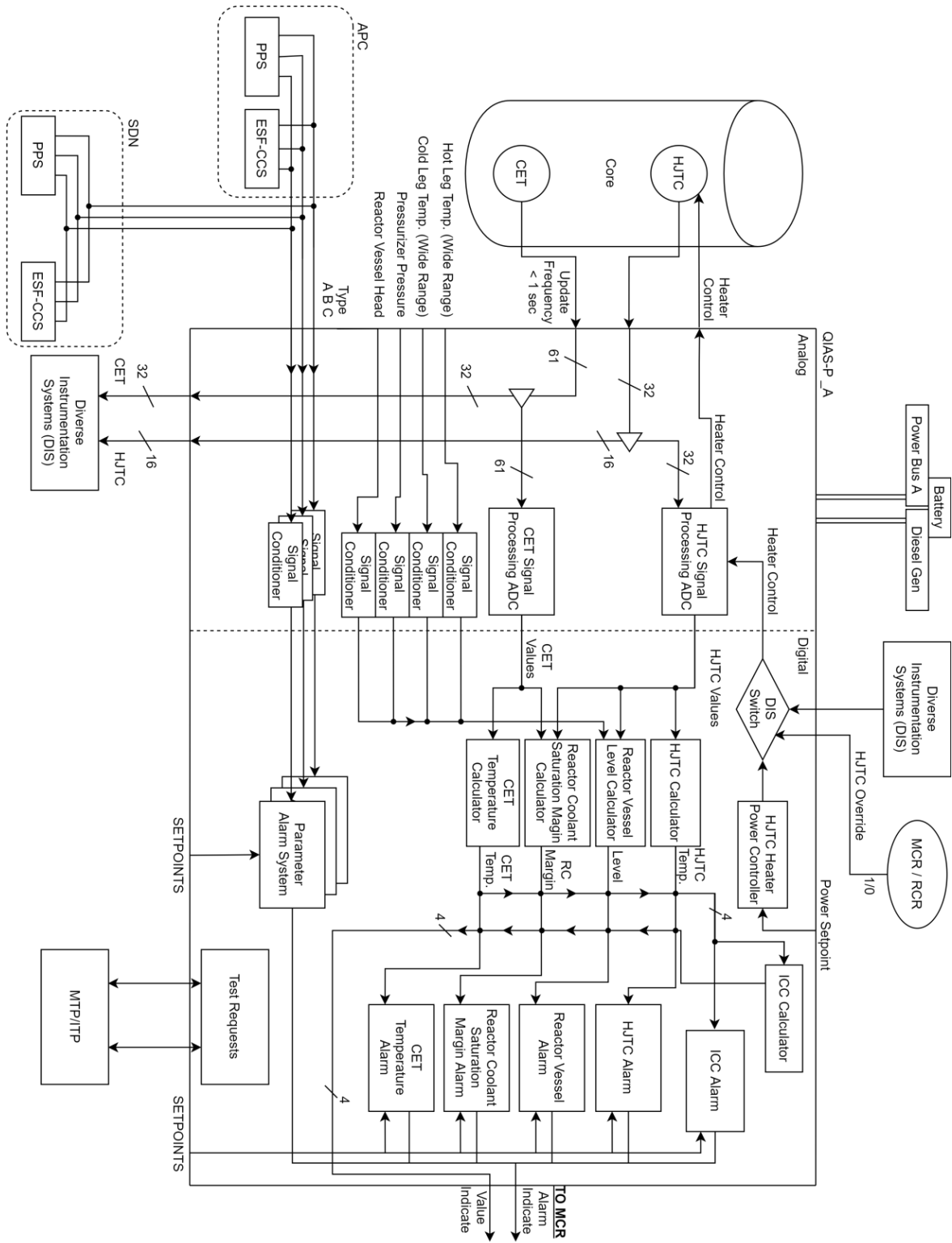

**Figure A1. QIAS-P Division A: Information system from core to main control room.**

# APPENDIX B

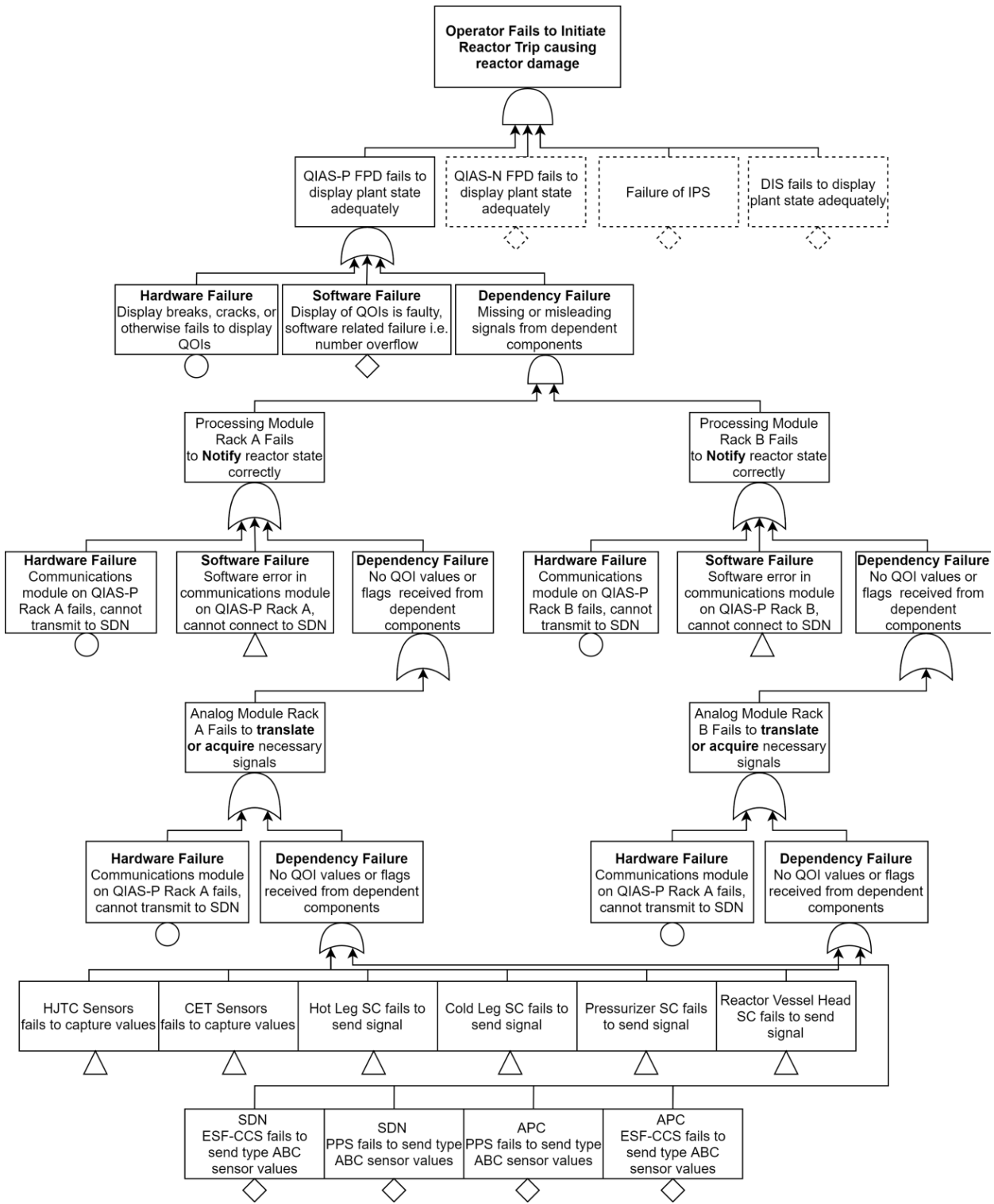

**Figure B1. QIAS-P system-level hardware fault tree with empty software failure branches.**